\documentclass[a4paper]{article}
\usepackage{amsmath}
\usepackage[affil-it]{authblk}

\begin{document}

\title{First order formalism for bimetric theories with connection interaction} 
\author{Idan Talshir
  \thanks{E-mail: \texttt{talshir@post.tau.ac.il}}}
\affil{School of Physics, Tel Aviv University, Ramat Aviv 69978, Israel}

\date{}

\maketitle
\begin{abstract}
We construct a first order Lagrangian formalism for bimetric theories with an interaction which is a general function of metrics and their derivatives, including non-analytic functions.  The first-order actions are fully equivalent to the original actions, with the same field equations for the metrics.
\end{abstract}

\section{Introduction}

First order formalism is a form of the action in which the field equations obtained are first order differential equations. There is an explicit first order formulation for general relativity with specific interactions \cite{MTW}, but a first order formalism for general interactions has not been constructed. Bimetric theories are gravity theories with two metrics and interaction between metrics and their derivatives. A first order formalism for general bimetric theories has not been previously studied. With the first order formalism, formulating a generalized Hamiltonian, identifying  constraints and calculating their algebra, are more easy and straightforward than in the standard form. In this paper we construct a first order formalism for general bimetric theories. The method we develop can be applied to general relativity.\\ 
  
 In general relativity a first order formalism is obtained when one refers to the metric field and the connections as independent in the variation, and formulating the Lagrangian so it is linear in metric derivatives. In a case where there is no interaction with other fields, it can be shown \cite{MTW} that from the equations derived by variations of the connections one can obtain the relations between the connections and the metric that defines the connections as Christoffel connections, and the field equations for the metric are identical to those obtained in the standard form (second order form), after substituting the solutions for the connections.\\

Interaction of gravity with a scalar field (mass) or electromagnetic field does not contain metric derivatives. In this case too the equations of motion obtained after transformation of the Einstein-Hilbert Lagrangian to first order form are still  equivalent to those obtained from the action in the standard form. This is because the variation of the interaction by the connections does not change the algebraic equations obtained from the free Lagrangian.\\
In cases where the connections explicitly appear in the interaction, as in covariant derivatives of spinor field, a variation by the connection does not yield the desired relations which define the connections as Christoffels. Also in modified gravity theories such as f(R), if we form Ricci's tensor as in the first order form of general relativity, we do not obtain the desired relations for the connections \cite{f(R)}. As far as we know, a first order formalism for f(R) that preserves the metric field equations has not yet been proposed.\\
Bimetic theories are gravity theories with two second rank tensor dynamical fields. These tensors are used to construct affine connections. The free action for each tensor is the Einstein-Hilbert action. Some of these theories have been proposed as a modification of general relativity designed to predict cosmological and astrophysical phenomena. In order to predict these phenomena within the framework of general relativity, a huge amount of dark matter and dark energy is required \cite{CaFra}.  Milgrom suggested ``Bimond''\cite{Milgrom1}- a theory with two metrics: One metric is coupled to matter, and so determines the motion of test particles, and the other, a ``twin metric'', is used to construct, with the first (ordinary) metric, an interaction term. That interaction term becomes dominant at low accelerations (the MOND regime), and the theory can produce flat rotation curves with less dark matter than General Relativity needs in order to explain the  phenomena. \\
 It should be noted that a large set of multiple-metric theories that can be approximated by the Pauli-Fierz action suffer from Boulware-Deser ghost instabilities \cite{inconsistencies}. A sub-set of these theories that avoid the ghost problem has been constructed \cite{RandH}, but recent developments \cite{SandA} cast doubt on the relevance of these. However, BiMOND gravity theories are outside the framework of this debate, since these theories are intended to give the MOND potential in the weak field limit (for the flat rotation curve phenomenon), and not the Newtonian potential, and therefore cannot be linearized near the double Minkowski metrics, and cannot be approximated by the Pauli-Fierz action.\\

 \section{Construction}
 We begin with a presentation of the general bimetric action, and a simple transformation to first order form in a special case where there are no metric derivatives (derivatives of metrics with respect to space-time coordinates) in the interaction. We continue by formulating the 
 requirements for a first order form equivalent to the standard (second order form), and develop in detail a method that fulfils these requirements in a square derivative interaction. We then generalize the method to any power of derivative, for analytical and a class of non-analytical interactions.
 We emphasize that this issue of the equivalent first order formulation has not been fully studied before, and there is no detailed construction in previous literature.\\
The general action for a bimetric theory can be written, as mentioned, in the form
 \begin{equation} 
 \label{twentyfive}
I = \int {{\cal L}{d^4}x = } \int {(\alpha {{\cal L}_G} + \beta {{\hat {\cal L}}_G} + {{\cal L}_I}){d^4}x} 
 \end{equation}
 
 where the scalar densities in the parentheses are the Einstein-Hilbert scalar density for the metric  
 ${g_{\mu \nu }}$, ${{\cal L}_G} =  - \frac{1}{{16\pi G}}R\sqrt { - g} $, a scalar density for the twin metric ${\hat g_{\mu \nu }}$, ${{\hat {\cal L}}_G} =  - \frac{1}{{16\pi G}}\hat R\sqrt { - \hat g} $, and an interaction term which depends on both metric and other fields and their derivatives, i.e., $ {{\cal L}_I}[{g_{\mu \nu }};{g_{\mu \nu ,\rho }};{{\hat g}_{\mu \nu }};{g_{\mu \nu ,\rho }};\psi ;{\psi _{,\rho }}]$, where $\psi $ symbolizes any other fields with any tensorial properties. The coefficients $\alpha $ and $\beta$ are any constants.  We notice that the derivatives of the metric (and other fields) are not tensors, but they can be arranged to construct a tensor, e.g. the difference between the Christoffel connections of the two metrics, and to construct a scalar density ${{\mathcal{L}}_{I}}$.\\
 If the interaction term ${\mathcal{L}_{\operatorname{int} }}$ does not explicitly depend on metric derivatives, the generalization  to first order form from general relativity is simple. Each ``kinetic'' term ${{\cal L}_G},{{\hat {\cal L}}_G}$ can be presented in first order form of general relativity:
\begin{equation}
\label{three}
{{{\tilde{\mathcal{L}}}}_{G}}=\sqrt{-g}{{g}^{\alpha \beta }}{{R}_{\alpha \beta }}[\Gamma _{\alpha \beta }^{\gamma }]\,\,\,,\,\,\,{{{\tilde{\hat{\mathcal{L}}}}}_{G}}=\sqrt{-\hat{g}}{{{\hat{g}}}^{\alpha \beta }}{{{\hat{R}}}_{\alpha \beta }}[\hat{\Gamma }_{\alpha \beta }^{\gamma }]
\end{equation}
The first order Lagrangian is obtained by modifying the kinetic terms ${{\mathcal{L}}_{G}},{{\hat{\mathcal{L}}}_{G}}$ in the action \eqref{twentyfive} to ${{\tilde{\mathcal{L}}}_{G}},{{\tilde{\hat{\mathcal{L}}}}_{G}}$ given above.
 The equations obtained by variation with respect to $ \Gamma _{\alpha \beta }^\gamma ,\hat \Gamma _{\alpha \beta }^\gamma $ yields the relations that define them as Christiffels for the metrics $ {g^{\alpha \beta }},{{\hat g}^{\alpha \beta }} $ respectively, and the metric field equations are preserved, i.e. the metric field equations one derives from the original bimetric action \eqref{twentyfive} are the same equations one derives from the first order action after substituting the solution for connections constraint equations.
   
\subsection{General considerations}
In order to construct a first order form for a given bimetric theory, we require that the interaction term  in the new form does not change the constraint equations which follow from the free action when varied with respect to connections, i.e.,
\begin{equation}
 \label{one}
\left( \frac{\delta {{{\tilde{\mathcal{L}}}}_{I}}}{\delta \Gamma _{\alpha \beta }^{\gamma }} \right)[\Gamma _{\alpha \beta }^{\gamma }=\{_{\alpha \beta }^{\,\,\,\gamma }\},\hat{\Gamma }_{\alpha \beta }^{\gamma }=\{_{\alpha \beta }^{\,\,\,\gamma }\hat{\}}]=\left( \frac{\delta {{{\tilde{\mathcal{L}}}}_{I}}}{\delta \hat{\Gamma }_{\alpha \beta }^{\gamma }} \right)[\Gamma _{\alpha \beta }^{\gamma }=\{_{\alpha \beta }^{\,\,\,\gamma }\},\hat{\Gamma }_{\alpha \beta }^{\gamma }=\{_{\alpha \beta }^{\,\,\,\gamma }\hat{\}}]=0
\end{equation}

where $ \{_{\alpha \beta }^{\,\,\,\gamma }\},\{_{\alpha \beta }^{\,\,\,\gamma }\hat{\}} $ are the Christoffel connections
\[\{_{\alpha \beta }^{\,\,\,\gamma }\}\equiv \frac{1}{2}{{g}^{\gamma \rho }}({{g}_{\rho \alpha ,\beta }}+{{g}_{\rho \beta ,\alpha }}-{{g}_{\alpha \beta ,\rho }})\,\,\,,\,\,\,\{_{\alpha \beta }^{\,\,\,\gamma }\hat{\}}\equiv \frac{1}{2}{{{\hat{g}}}^{\gamma \rho }}({{{\hat{g}}}_{\rho \alpha ,\beta }}+{{{\hat{g}}}_{\rho \beta ,\alpha }}-{{{\hat{g}}}_{\alpha \beta ,\rho }})\]
We also require that the field equations for the metrics are not changed as a result of the new formation of the interaction term:
\begin{equation}
\begin{gathered}\left({\frac{{\delta{{\tilde{\mathcal{L}}}_{I}}}}{{\delta{g^{\alpha\beta}}}}}\right)[\Gamma_{\alpha\beta}^{\gamma}=\{_{\alpha\beta}^{{\kern1pt }{\kern1pt }{\kern1pt }\gamma}\},\hat{\Gamma}_{\alpha\beta}^{\gamma}=\{_{\alpha\beta}^{{\kern1pt }{\kern1pt }{\kern1pt }\gamma}\hat{\}}]=\frac{{\delta{\mathcal{L}_{I}}}}{{\delta{g^{\alpha\beta}}}}\hfill\\
\left({\frac{{\delta{{\tilde{\mathcal{L}}}_{I}}}}{{\delta{{\hat{g}}^{\alpha\beta}}}}}\right)[\Gamma_{\alpha\beta}^{\gamma}=\{_{\alpha\beta}^{{\kern1pt }{\kern1pt }{\kern1pt }\gamma}\},\hat{\Gamma}_{\alpha\beta}^{\gamma}=\{_{\alpha\beta}^{{\kern1pt }{\kern1pt }{\kern1pt }\gamma}\hat{\}}]=\frac{{\delta{\mathcal{L}_{I}}}}{{\delta{{\hat{g}}^{\alpha\beta}}}}\hfill
\end{gathered}
\label{one42}
\end{equation}
where the left hand sides of equations \eqref{one42} correspond to variation of the modified interaction with respect to the metrics and substituting the solution for the connections as Christoffel connections. The above equation is maintained when the modified interaction is equal to the original interaction after substituting the solution of the connections as Christoffels:, %as can be seen by the equality:
\begin{equation}
\begin{gathered}\frac{{\delta{\mathcal{L}_{I}}}}{{\delta{g^{\alpha\beta}}}}=\frac{{\delta{{\tilde{\mathcal{L}}}_{I}}[\Gamma_{\alpha\beta}^{\gamma}=\{_{\alpha\beta}^{{\kern1pt }{\kern1pt }{\kern1pt }\gamma}\},\hat{\Gamma}_{\alpha\beta}^{\gamma}=\{_{\alpha\beta}^{{\kern1pt }{\kern1pt }{\kern1pt }\gamma}\hat{\}}]}}{{\delta{g^{\alpha\beta}}}}=\hfill\\
=\left({\frac{{\delta{{\tilde{\mathcal{L}}}_{I}}}}{{\delta{g^{\alpha\beta}}}}}\right)[\Gamma_{\alpha\beta}^{\gamma}=\{_{\alpha\beta}^{{\kern1pt }{\kern1pt }{\kern1pt }\gamma}\},\hat{\Gamma}_{\alpha\beta}^{\gamma}=\{_{\alpha\beta}^{{\kern1pt }{\kern1pt }{\kern1pt }\gamma}\hat{\}}]+\frac{{\delta{{\tilde{\mathcal{L}}}_{I}}}}{{\delta\Gamma_{\alpha\beta}^{\gamma}}}\frac{{\delta\{_{\alpha\beta}^{{\kern1pt }{\kern1pt }{\kern1pt }\gamma}\}}}{{\delta{g^{\alpha\beta}}}}=\left({\frac{{\delta{{\tilde{\mathcal{L}}}_{I}}}}{{\delta{g^{\alpha\beta}}}}}\right)[\Gamma_{\alpha\beta}^{\gamma}=\{_{\alpha\beta}^{{\kern1pt }{\kern1pt }{\kern1pt }\gamma}\},\hat{\Gamma}_{\alpha\beta}^{\gamma}=\{_{\alpha\beta}^{{\kern1pt }{\kern1pt }{\kern1pt }\gamma}\hat{\}}]\hfill
\end{gathered}
\label{elevenone}
\end{equation}
where the last equality is obtained by using \eqref{one}.

\subsection{General interaction model of bimetric theories}
As a first step in the construction of a first order formalism for general bimetric theories, we want to present the interaction as a combination of terms, each composed of a part that depends only on connections, and a part that is a function of metrics without their derivatives.\\
The interaction term is a scalar density, and can be presented as
\[{{\mathcal{L}}_{I}}=F(S)D\]
where $S$ is a dimensionless scalar, and $D$ is a scalar density which depends on determinants of the the metrics and absorbs a constant with the appropriate lengh dimensions.\\
From the two metrics one can generate a non trivial tensor that involves only first order derivatives, that is the difference of their Levi-Civita connections.
\[C_{\alpha \beta }^{\gamma }\equiv \Gamma _{\alpha \beta }^{\gamma }-\hat{\Gamma }_{\alpha \beta }^{\gamma }\]
By using this tensor, and metrics and their determinants, various scalars can be constructed, e.g.
\[C_{\alpha \beta }^\gamma C_{\mu \nu }^\rho {g^{\alpha \beta }}{g^{\mu \nu }}{{\hat g}_{\alpha \beta }}\,\,\,,\,\,\,C_{\alpha \beta }^\gamma C_{\mu \nu }^\mu {g^{a\nu }}\delta _\gamma ^\alpha \]
We notice that any scalar that is constructed in this way can be expressed as an outer product of an even number $N$ of $C_{\alpha \beta }^\gamma $ tensors contracted with a tensor with $N$ upper indexes and $2N$ lower indexes. The last tensor has no explicit dependence on the connections. For example: $C_{\alpha \beta }^\gamma C_{\mu \nu }^\mu {g^{a\nu }}\delta _\gamma ^\alpha  = C_{\alpha \beta }^\gamma C_{\mu \nu }^\rho A_{\gamma \rho }^{\alpha \beta \mu \nu }$ 
where $A_{\gamma \rho }^{\alpha \beta \mu \nu } \equiv \delta _\rho ^\mu {g^{a\nu }}\delta _\gamma ^\alpha $\\
If we restrict the theories to derivative terms squared, there are just two tensors $C_{\alpha \beta }^\gamma $
in such a scalar.
\subsection{First order for square derivatives bimetric theories}
With the assumption that
\[F(S) = S\,\,\,,\,\,\,\,\,S = C_{\alpha \beta }^\gamma C_{\mu \nu }^\rho A_{\gamma \rho }^{\alpha \beta \mu \nu }\]
we want to transform ${{\mathcal{L}}_{I}}$ into the form ${{{\tilde{\mathcal{L}}}}_{I}}$ that satisfies \eqref{one} and \eqref{one42}.\\
We construct a model which is linear in metric derivatives.
\begin{equation}
\label{eleven}
{{{\tilde{\mathcal{L}}}}_{I}}=\left( \tilde{C}_{\alpha \beta }^{\gamma }C_{\mu \nu }^{\rho }+C_{\alpha \beta }^{\gamma }\tilde{C}_{\mu \nu }^{\rho }-\tilde{C}_{\alpha \beta }^{\gamma }\tilde{C}_{\mu \nu }^{\rho } \right)A_{\gamma \rho }^{\alpha \beta \mu \nu }D
\end{equation}
where $ \tilde C_{\alpha \beta }^\gamma ,\tilde C_{\mu \nu }^\rho $ are expressed explicitly as functions of the connections only, without the metrics and their derivatives, and $C_{\alpha \beta }^\gamma ,C_{\mu \nu }^\rho $ are presented explicitly in the standard form
\[C_{\alpha \beta }^\gamma  = \{ _{\alpha \beta }^\gamma \}  - \{ _{\alpha \beta }^\gamma \hat \} \]
as functions of the metrics and their derivatives.\\
Variation with respect to connections results in
\begin{equation}
\begin{gathered}\frac{{\delta{{\tilde{\mathcal{L}}}_{I}}}}{{\delta\Gamma_{\alpha\beta}^{\gamma}}}=\frac{{\delta\tilde{C}_{\alpha\beta}^{\gamma}}}{{\delta\Gamma_{\alpha\beta}^{\gamma}}}\left({C_{\mu\nu}^{\rho}-\tilde{C}_{\mu\nu}^{\rho}}\right)+\frac{{\delta\tilde{C}_{\mu\nu}^{\rho}}}{{\delta\Gamma_{\alpha\beta}^{\gamma}}}\left({C_{\alpha\beta}^{\gamma}-\tilde{C}_{\alpha\beta}^{\gamma}}\right)\hfill\\
\frac{{\delta{{\tilde{\mathcal{L}}}_{I}}}}{{\delta\hat{\Gamma}_{\alpha\beta}^{\gamma}}}=\frac{{\delta\tilde{C}_{\alpha\beta}^{\gamma}}}{{\delta\hat{\Gamma}_{\alpha\beta}^{\gamma}}}\left({C_{\mu\nu}^{\rho}-\tilde{C}_{\mu\nu}^{\rho}}\right)+\frac{{\delta\tilde{C}_{\mu\nu}^{\rho}}}{{\delta\hat{\Gamma}_{\alpha\beta}^{\gamma}}}\left({C_{\alpha\beta}^{\gamma}-\tilde{C}_{\alpha\beta}^{\gamma}}\right)\hfill
\end{gathered}
\label{four}
\end{equation}
so the solution 
\begin{equation}
\label{five}
\Gamma _{\alpha \beta }^{\gamma }=\{_{\alpha \beta }^{\,\,\,\gamma }\},\hat{\Gamma }_{\alpha \beta }^{\gamma }=\{_{\alpha \beta }^{\,\,\,\gamma }\hat{\}}
\end{equation}
sets the right hand side of \eqref{four} to zero, and is consistent with the equations
\[\frac{{\delta \mathcal{L}}}{{\delta \Gamma _{\alpha \beta }^\gamma }} = 0\,\,\,,\,\,\,\frac{{\delta \mathcal{L}}}{{\delta \hat \Gamma _{\alpha \beta }^\gamma }} = 0\]
where $\mathcal{L}$ is the whole Lagrangian, and ${{\mathcal{L}}_{G}},{{\hat{\mathcal{L}}}_{G}}$ are given in the form of \eqref{three}.
After substituting the solution \eqref{five} the original and the transformed Lagrangian are equal:
\begin{equation}
{{{\tilde{\mathcal{L}}}}_{I}}\left[ {{g}^{\alpha \beta }},{{g}^{\alpha \beta }}_{,\rho },{{{\hat{g}}}^{\alpha \beta }},{{{\hat{g}}}^{\alpha \beta }}_{,\rho },\Gamma _{\alpha \beta }^{\gamma }=\{_{\alpha \beta }^{\gamma }\},\hat{\Gamma }_{\alpha \beta }^{\gamma }=\{_{\alpha \beta }^{\gamma }\hat{\}} \right]={{\mathcal{L}}_{I}}\left[ {{g}^{\alpha \beta }},g_{,\rho }^{\alpha \beta },{{{\hat{g}}}^{\alpha \beta }},{{{\hat{g}}}^{\alpha \beta }}_{,\rho } \right]
\end{equation}
so the field equations for the metrics are preserved.
\subsection{First order formalism for bimetric theories with a product of $N$ derivatives}
We assume $F(S) = S$ where
\[S = \left( {\prod\limits_{i = 1}^N {C_{\alpha i\beta i}^{\gamma i}} } \right)A_{\alpha 1\beta 1...\alpha N\beta N}^{\gamma 1...\gamma N}\]
We then generalize the result \eqref{eleven} for square derivatives to:
\begin{equation}
\label{twelve}
{{\tilde{\mathcal{L}}}_{I}} = \left( {\sum\limits_{j = 1}^N {C_{\alpha j\beta j}^{\gamma j}} \prod\limits_{i \ne j} {\tilde C_{\alpha i\beta i}^{\gamma i}}  + (1 - N)\prod\limits_{i = 1}^N {\tilde C_{\alpha i\beta i}^{\gamma i}} } \right)A_{\alpha 1\beta 1...\alpha N\beta N}^{\gamma 1...\gamma N}D
\end{equation}
where $C_{\alpha i\beta i}^{\gamma i}$ functionally depend only on metrics and their derivatives, and $\tilde C_{\alpha i\beta i}^{\gamma i}$ functionally depend only on connections.\\
Variation of \eqref{twelve} with respect to connections gives:
\begin{equation}
\begin{gathered}\frac{{\delta{{\tilde{\mathcal{L}}}_{I}}}}{{\delta\Gamma_{\alpha\beta}^{\gamma}}}=\left({\sum\limits _{j=1}^{N}{\left({C_{\alpha j\beta j}^{\gamma j}-\tilde{C}_{\alpha j\beta j}^{\gamma j}}\right)\left({\sum\limits _{k\ne j}^{N}{\frac{{\delta\tilde{C}_{\alpha k\beta k}^{\gamma k}}}{{\delta\Gamma_{\alpha\beta}^{\gamma}}}}\prod\limits _{i\ne j,k}{\tilde{C}_{\alpha i\beta i}^{\gamma i}}}\right)}}\right)A_{\alpha1\beta1...\alpha N\beta N}^{\gamma1...\gamma N}D\hfill\\
\frac{{\delta{{\tilde{\mathcal{L}}}_{I}}}}{{\delta\hat{\Gamma}_{\alpha\beta}^{\gamma}}}=\left({\sum\limits _{j=1}^{N}{\left({C_{\alpha j\beta j}^{\gamma j}-\tilde{C}_{\alpha j\beta j}^{\gamma j}}\right)\left({\sum\limits _{k\ne j}^{N}{\frac{{\delta\tilde{C}_{\alpha k\beta k}^{\gamma k}}}{{\delta\hat{\Gamma}_{\alpha\beta}^{\gamma}}}}\prod\limits _{i\ne j,k}{\tilde{C}_{\alpha i\beta i}^{\gamma i}}}\right)}}\right)A_{\alpha1\beta1...\alpha N\beta N}^{\gamma1...\gamma N}D\hfill
\end{gathered}
\label{44three}
\end{equation}
where we have used the identity \[\sum\limits_{k=1}^{N}{\delta \tilde{C}_{\alpha k\beta k}^{\gamma k}}\prod\limits_{i\ne k}{\tilde{C}_{\alpha i\beta i}^{\gamma i}}=\frac{1}{N-1}\sum\limits_{j=1}^{N}{\tilde{C}_{\alpha j\beta j}^{\gamma j}\sum\limits_{k\ne j}^{N}{\delta \tilde{C}_{\alpha k\beta k}^{\gamma k}}\prod\limits_{i\ne j,k}{\tilde{C}_{\alpha i\beta i}^{\gamma i}}}\] for the variations of the second addend in \eqref{twelve}. The definition of the connections  as Levi-Civita (Chiristoffel) $\Gamma _{\alpha \beta }^\gamma ,\hat \Gamma _{\alpha \beta }^\gamma $ is then a solution for the field equations, and
\[\tilde{\mathcal{L}}\left[ \Gamma _{\alpha \beta }^{\gamma }=\{_{\alpha \beta }^{\gamma }\},\hat{\Gamma }_{\alpha \beta }^{\gamma }=\{_{\alpha \beta }^{\gamma }\hat{\}} \right]=\mathcal{L}\]
so the field equations for the metrics are preserved.
\subsection{First order formalism for an analytic interaction}

Assuming 
\[{\mathcal{L}_I} = F(S)D\]
and $F(S)$ analytic at $S=0$ the interaction can be developed as
\[{\mathcal{L}_I} = \sum {{a_n}{S^n}} D\]
We represent ${{S}^{n}}$ as a contraction of two $3nN$ tensors
\[{{S}^{n}}={{\left( \left( \prod\limits_{i=1}^{N}{C_{\alpha i\beta i}^{\gamma i}} \right)A_{\alpha 1\beta 1...\alpha N\beta N}^{\gamma 1...\gamma N} \right)}^{n}}=\left( \prod\limits_{l=1}^{n}{\prod\limits_{i=1}^{N}{C_{\alpha i\beta i}^{\gamma i}}} \right)\prod\limits_{l=1}^{n}{A_{\alpha l1\beta l1...\alpha lN\beta lN}^{\gamma l1...\gamma lN}}\]
and use it to generalise the result \eqref{twelve} we obtained for $S$ to ${{S}^{n}}$, i.e., 

\begin{equation}
\label{44two}
{{{\tilde{S}}}^{n}}={{\left( A_{\alpha 1\beta 1...\alpha N\beta N}^{\gamma 1...\gamma N}\prod\limits_{i=1}^{N}{\tilde{C}_{\alpha i\beta i}^{\gamma i}} \right)}^{n-1}}\left( n\sum\limits_{j=1}^{N}{C_{\alpha j\beta j}^{\gamma j}}\prod\limits_{i\ne j}{\tilde{C}_{\alpha i\beta i}^{\gamma i}}+(1-nN)\prod\limits_{i=1}^{N}{\tilde{C}_{\alpha i\beta i}^{\gamma i}} \right)A_{\alpha 1\beta 1...\alpha N\beta N}^{\gamma 1...\gamma N}D
\end{equation}
We now modify the interaction 
 ${{\mathcal{L}}_{I}}$ to
\begin{equation}
\label{eleventwo}
{{\tilde{\mathcal{L}}}_{I}}=\sum{{{a}_{n}}{{{\tilde{S}}}^{n}}}D
\end{equation}
The modified interaction ${\mathcal{L}_I}$ is linear in metric derivatives.
Variation of \eqref{eleventwo} with respect to the connections gives a linear combination of equations in the form of \eqref{44three} so the modified interaction keeps the definition of the connections as Christoffels. The modified interaction ${\mathcal{L}_I}$ is equal to the original interaction after substituting the solution of the connections as Christoffels, so it preserves the field equations for the metrics by \eqref{elevenone}.

\subsection{First order formalism for non-analytic interaction}
Suppose $F(S)$ is not analytic at $S=0$ but an analytic function $H(S)$ can be defined so that
\[F(S) = H(S){S^{ - b}}\]
where $b$ is some real number, then $H(S) = {a_n}{S^n}$ and
\begin{equation}
\label{44one}
  F(S) = {a_n}{S^{n - b}}  
\end{equation}

So one can develop the interaction in the form 
\begin{equation}
\label{44five}
{{\mathcal{L}}_{I}}=\sum{{{a}_{n}}{{S}^{n-b}}D}
\end{equation} 
We constructed the modified scalar ${{{\tilde{S}}}^{n}}$ by representing ${{S}^{n}}$ as a contraction of an  outer product of integer n $C_{\alpha i\beta i}^{\gamma i}$ and $A_{\alpha 1\beta 1...\alpha N\beta N}^{\gamma 1...\gamma N}$ tensors, but the result \eqref{44two} is valid also for negative and non-integer powers of $S$. To show this, we vary ${{{\tilde{S}}}^{n}}$ with respect to the connections:
\begin{equation}
\frac{\delta {{{\tilde{S}}}^{n}}}{\delta \Gamma _{\alpha \beta }^{\gamma }}={{\left( S[\tilde{C}] \right)}^{n-2}}\left( ({{n}^{2}}-n)\frac{\delta S[\tilde{C}]}{\delta \Gamma _{\alpha \beta }^{\gamma }}(\tilde{S}-S)+nS[\tilde{C}]\frac{\delta {{{\tilde{S}}}^{n}}}{\delta \Gamma _{\alpha \beta }^{\gamma }} \right)
\end{equation}
where $S[\tilde{C}]\equiv \left( \prod\limits_{i=1}^{N}{\tilde{C}_{\alpha i\beta i}^{\gamma i}} \right)A_{\alpha 1\beta 1...\alpha N\beta N}^{\gamma 1...\gamma N}$. The second term on the right hand side is zero when $\frac{\delta \tilde{S}}{\delta \Gamma _{\alpha \beta }^{\gamma }}=0$ which according to \eqref{44three} implies that the connections are Christoffel connections, and this is consistent with $\tilde{S}-S=0$ which sets the first addend to zero. This line of reasoning does not depend on $n$, and the generalization is thus justified.\\
We therefore modify the interaction \eqref{44five} to
\begin{equation}
{{{\tilde{\mathcal{L}}}}_{I}}=\sum{{{a}_{n}}{{{\tilde{S}}}^{n-b}}D}
\end{equation}
 where
\begin{equation}
\begin{gathered}{{\tilde{S}}^{n-b}}=\hfill\\
{\left({A_{\alpha1\beta1...\alpha N\beta N}^{\gamma1...\gamma N}\prod\limits _{i=1}^{N}{C_{\alpha i\beta i}^{\gamma i}}}\right)^{n-b-1}}\left({(n-b)\sum\limits _{j=1}^{N}{C_{\alpha j\beta j}^{\gamma j}}\prod\limits _{i\ne j}{\tilde{C}_{\alpha j\beta j}^{\gamma j}}+(1-N(n-b))\prod\limits _{i=1}^{N}{\tilde{C}_{\alpha i\beta i}^{\gamma i}}}\right)A_{\alpha1\beta1...\alpha N\beta N}^{\gamma1...\gamma N}D\hfill
\end{gathered}
\end{equation}
\subsection{Example: First order formalism for BIMOND}
The interaction in BIMOND is given in its general form by \cite{Milgrom1}:
\[{\mathcal{L}_I} =  - 2{\left( {g\hat g} \right)^{1/4}}f\left( {\frac{{{g^{1/4}}}}{{{{\hat g}^{1/4}}}}} \right)a_0^2M\left( {\frac{Y}{{a_0^2}}} \right)\]
where $Y$ is a scalar that is quadratic in the tensors $C_{\alpha \beta }^\gamma $. We define $S \equiv \left( {\frac{Y}{{a_0^2}}} \right)$. We require that $M'(S) \approx {S^{ - 1/4}}$ when $S \ll 1$ in order to obtain the MOND correction for the regime of acceleration that are approximately equal or smaller than the MOND acceleration ${a_0}$.
We define
\[H(S) = M(S){S^{1/4}}\]
so $H(S)$ analytic at $S=0$ and can be presented as $H(S) = \sum {{b_n}{S^n}} $ . We transform the interaction to:
\[{{{\tilde{\mathcal{L}}}}_{I}}=-2{{\left( g\hat{g} \right)}^{1/4}}f\left( \frac{{{g}^{1/4}}}{{{{\hat{g}}}^{1/4}}} \right)a_{0}^{2}\sum{{{b}_{n}}{{{\tilde{S}}}^{n-\tfrac{1}{4}}}}\]
where
\[{{\tilde S}^{n - \tfrac{1}{4}}} = {\left( {\tilde C_{\alpha '\beta '}^{\gamma '}\tilde C_{\mu '\nu '}^{\rho '}A_{\gamma '\rho '}^{\alpha '\beta '\mu '\nu '}} \right)^{n - 1\tfrac{1}{4}}}\left( {(n - \tfrac{1}{4})\left( {C_{\alpha \beta }^\gamma \tilde C_{\mu \nu }^\rho  + \tilde C_{\alpha \beta }^\gamma C_{\mu \nu }^\rho } \right) + (1 - 2(n - \tfrac{1}{4}))\tilde C_{\alpha \beta }^\gamma \tilde C_{\mu \nu }^\rho } \right)A_{\gamma \rho }^{\alpha \beta \mu \nu }\]
With this transformation of the interaction term ${{\mathcal{L}}_{I}}\to {{\tilde{\mathcal{L}}}_{I}}$ and the transformation \eqref{three} of kinetic terms, the BiMOND action is in first order form equivalent to the original form.

\section{Summary and discussion}
In this  paper we have constructed  a first order formalism for bimetric theories with interactions which are non-linear in metric derivatives. The model that is obtained is linear in metric derivatives, and has the same metric equations of motion as those obtained from the original bimetric model.\\
We developed the model for interaction with derivative terms squared, for derivative term in any power, and for analytic and non analytic interactions such as BiMOND interaction.\\
It should be noted that although we assumed that the non-trivial tensor ${C_{\alpha \beta }^\gamma }$ that can be composed from the metrics has some specific functional dependence on the connections, we have not used this dependence explicitly in the process of construction and proof. Therefore, the constructions that are proposed in this paper can fit other interactions which depend on connections, such as spinor interactions.\\
Transformation of the Lagrangian to first order formalism is a first step towards an Hamiltonian for any gauge theory, and specifically for gravity theories.\cite{ADM} \cite{Faddeev}.(see \cite{Talshir} for generalization to specific bimetric theories with no dependence on connections). These theories are constrained (as all gauge theories are \cite{TH}), and there is a general procedure for constructing an Hamiltonian for such systems introduced by Dirac \cite{Dirac2}. Dirac applied his method to general relativity, with no metric derivative in the interaction \cite{Dirac1}. However, solving the constraint equations may be a difficult task, and a generalized Hamiltonian is sufficient for some applications \cite{FandP}. The first order formalism may simplify the identification of the generalized Hamiltonian. From first order form of the Lagrangian, one can determine the algebra of first class constraints \cite{Teitelboim}.\\
The method that is presented in the paper may be implemented in a general relativistic model with interactions that contain metric derivatives, like interactions of gravity with a spinor field, where the covariant derivatives are in explicit form. 

\subsection*{Acknowledgements}
I want to thank Lawrence P. Horwitz for his careful reading and useful discussions.
I gratefully acknowledge financial support from Ariel University.

\end{document}